\documentclass[preprint]{article}
\usepackage{graphicx}
\usepackage{authblk}
\usepackage{latexsym}

\begin{document}

\title{Random matrix analysis for gene interaction networks in cancer cells}

\author[]{Ayumi Kikkawa}
\affil{Mathematical and Theoretical Physics Unit, \\
Okinawa Institute of Science and Technology Graduate University,\\
1919-1 Tancha, Onna-son, Kunigami-gun, Okinawa, 904-0495 Japan \\
e-mail: akikkawa@oist.jp \\}




\maketitle

\begin{abstract}
Investigations of topological uniqueness of gene interaction networks in cancer cells are essential for understanding the disease. Although cancer is considered to originate from the topological alteration of a huge molecular interaction network in cellular systems, the theoretical study to investigate such complex networks is still insufficient.  It is necessary to predict the behavior of a huge complex interaction network from the behavior of a finite size network.  Based on the random matrix theory, we study the distribution of the nearest neighbor level spacings $P(s)$ of interaction matrices of gene networks in human cancer cells. The interaction matrices are computed using the Cancer Network Galaxy (TCNG) database which is a repository of gene interactions inferred by a Bayesian network model.  $256$ NCBI GEO entries regarding gene expressions in human cancer cells have been used for the inference.  We observe the Wigner distribution of $P(s)$ when the gene networks are dense networks that have more than $\sim 38,000$ edges.   In the opposite case, when the networks have smaller numbers of edges,  the distribution $P(s)$ becomes the Poisson distribution.  We investigate relevance of  $P(s)$ both to the sparseness of the networks and to edge frequency factor which is the reliance (likelihood) of the inferred gene interactions. 
\end{abstract}

\section{Introduction}
\label{intro}
      
   There have been many experimental or theoretical studies of molecular interaction networks in cancer cells.
   They revealed gradually that cancer cells are characterized by alterations of the intermolecular networks.
   By observing the gene co-expression patterns with high-throughput experiments such as microarrays or next-generation
   sequencing technologies, we can study gene interaction networks related to cancer. \cite{Peer2011}

   Behaviors of the complex gene networks are unknown totally even in normal cells. 
   Recent studies have revealed that there are large regions in DNA that do
   not code any protein, although they are highly transcribed.
   These transcripts are called non-coding RNA. \cite{Mattick2006}
   The importance of such transcripts as regulators of the gene
   expressions has become widely known to date from various
   experiments. \cite{Flamant2010, Guo2013}  It is suggested that the
   non-coding RNA bind to other transcripts selectively
   and thus regulate the gene expressions. \cite{Leveille2015, Taft2010}
   Micro RNA, which are about 20nt subsets of the non-coding RNA,
   have also been observed negatively regulating the gene expressions through
   interactions with other RNA or even with DNA. \cite{Bartel2009}
   These  interaction networks of various transcripts have important role in cellular cycles including
   cell development, proliferation, apoptosis and disease. \cite{Cech2014}
   For example, several micro RNA behave as inhibitors for specific
   interactions in the gene network, and they act as potential
   oncogenes or tumor genes by permitting uncontrolled
   proliferation of damaged cells. \cite{Lim2015}
   
   Relations between human diseases and modifications of the interacting molecular networks have also been 
   extensively studied theoretically. \cite{Creixell2015, JonBates2006, Jonsson2006, Pamplona2012, Gulati2013, Kling2015}
   Such investigations in cancer cells are very important to discover new biomarkers or to classify the symptoms in detail. \cite{ Goh2007, Vidal2011, Rai2017}
   
   The high-throughput experiments provide huge data of molecular interaction networks, in which 20,000 to two million elements are involved within a single assay.
   Such experiments became very popular owing to the wide distributions of commercial platforms.
   Moreover, these expression data are accessible on the internet. For example the NCBI GEO (http://www.ncbi.nlm.nih.gov/geo/) provides a public database of gene expression data. \cite{Barrett2011}
   By using the database, it is even possible to perform a meta-analytical study of gene expressions in cancer cells.
   Since cancer is characterized by complex topological modulations of a huge interaction network of various transcripts, the theoretical study to investigate huge networks is necessary.
   It is also important to discuss how to predict the behavior of the large network limit from the behavior of the finite size networks obtained from experiments.

   Computational inference of gene interactions from the
      expression data involves statistical methods such as clustering or principal component analysis.  
      Stochastic procedures are inevitable due to experimental noises.
      Individual interactions between genes are numerically
      calculated using algorithms based on the probabilistic graphical models.
      The Markov network or Bayesian network models are the main frameworks in the study of
      gene network classifications, 
      and there are a lot of studies on gene regulatory networks, protein-protein networks      
      and on molecular pathways in a variety of organisms.
      \cite{Friedman2000, Friedman2004, Marbach2012, Segal2003,  Schafer2005, Zitnik2015} 
      
      There are several public gene interaction network databases
      which are based on, for example, the mutual information
      (ARACNE) \cite{Margolin2006}, the Bayesian network approach 
      (SiGN-BN) \cite{Tamada2011} and much more. \cite{Novere2015, Creixell2015}
      In such databases, the inferred interactions are characterized by confidence
      or likelihood factors which evaluate certainties of the interactions.  
      The key point of learning the gene network classification is how to improve the choice of the  likelihood 
      factors by integrating related informations of the cells.
      
    Also, network inference algorithms usually require highly time consuming calculations since they involve
    huge iterative learning process.
    Thus the inferred network size is rather small.   On the other hand, for the investigations of 
    the disease such as cancer, a macroscopic view of the huge complex network topology is necessary.
    We have to balance the amount of computing resources
    and the choice of adequate thresholds of the likelihood factors in various aspects through the computations.  
    Some estimations of how variations of the thresholds are related to a topological modification of the large networks are necessary.  
   
    In this paper, we discuss how sparseness of gene networks, thresholds of likelihood factors of edges
    and sample sizes in expression data are related to the changes in global topologies 
    of the interacting gene networks by using the method of the random matrix theory.   We also discuss
    possibilities of improving the huge network inference algorithms with this method.

    The random matrix theory (RMT)\cite{Mehta1991, Akemann2011} has been applied to a variety of
    fields not only in physics but also in biology.
    There are several studies in which RMT is applied in the analysis of complex
    networks including protein-protein interaction networks and gene co-expressions. \cite{Luo2006a, Luo2006b, Luo2007, Rai2014, Agrawal2014, Jalan2012}
    We have also studied protein-protein interaction networks in many
    organisms such as human, yeast, and Arabidopsis with the random matrix theory
    and obtained a universal (system independent) behavior of the distribution of the nearest
    neighbor level (NNL) spacings $P(s)$ of interaction matrices.  The NNL spacings $s$
    are the spacings between two adjacent eigenvalues of gene (or protein-protein) interaction matrices.
    The universal distribution $P(s)$ in large matrix size $N$ is called the Wigner distribution. 
    From these studies we consider that RMT gives a clue to the analysis of the huge gene interaction networks in living cells. 
    
    In the random matrix theory, it is necessary to take ensemble average to evaluate statistics of the eigenvalues.  
    In the gene expression experiments, the number of expressing genes is huge, but the number of samples is limited.  It is not self-evident whether the Wigner's surmise (the Wigner distribution of $P(s)$) is correct in the gene interaction matrices which are inferred from such data. 
    Also the interaction matrices of biological networks become sparse matrices in many cases.  
    In the sparse random matrices, the number of nonzero elements in each row (the degree $k$ of each node) takes a finite value in the thermodynamic limit.  
    It has been shown theoretically that the eigenvalue distribution of the sparse random matrix has a special behavior at the center part and the tails of the Wigner's semi-circle.
    In this work, the level spacing distribution behavior in the large $N$ limit is numerically examined from a finite number of eigenvalues of the gene interaction matrices using so-called unfolding method which is described in section 2.4. \cite{Reichl1992}  

    We use the gene interaction data  from the Cancer Network Galaxy (TCNG) database, where
    the gene interactions were computationally inferred with the non-parametric Bayesian algorithm named SiGN-BN. \cite{Tamada2011}
    Gene expression data from various cancer cells are used for the Bayesian network calculations in TCNG.
    In this database, each inferred gene interactions (directed edges) takes a factor called the edge frequency, which indicates the reliability (or the likelihood) of the gene interaction.
We study distribution of NNL spacings $P(s)$ in each of 256 gene networks in TCNG database, and investigate how the large $N$ limit behavior is altered due to various network attributes.

\section{Method}
\label{method}

\subsection{The Cancer Network Galaxy (TCNG) database}
\label{sec:2.1}
The Cancer Network Galaxy (TCNG) (http://tcng.hgc.jp) is the database of
computationally inferred gene interaction networks from the NCBI GEO
expression data that are related to human cancer samples. 
256 GEO entries are selected for the gene interaction inference calculation based on the Bayesian network 
model.  
TCNG (Release 0.14 built on Wed Mar 30 15:00:31 2016) currently stores
more than 16 million edges (interactions) between 24,907 nodes (genes). The edges are given
with directions and the edge frequency factors as their edge attributes. 
Learning of Bayesian networks are heavily time and memory consuming computations.  
With the use of the algorithm named NNSR (the neighbor node sampling and
repeat), they have obtained considerably large gene interaction networks
using the RIKEN AICS K supercomputer.\cite{Tamada2011}

In the Bayesian network model, directed edges connecting two nodes
express  causal relationships between them.  In the case of  the gene interaction networks, 
the directions of edges may infer regulatory relationships between genes.  
We study the case that the gene interaction matrices are real symmetric by neglecting the directions of the edges.
We thus study the Gaussian orthogonal ensemble (GOE) RMT.
In the real biological systems, where the directionality of the molecular interactions plays an important role, the Gaussian unitary ensemble (GUE) RMT may also be studied.
Number of studies show that the universality of $P(s)$ is independent of the details of the systems to be investigated, and we think that it is important to investigate whether such universality can be observed also in the undirected gene interaction networks.

\subsection{The random matrix theory}
\label{sec:2.2}
   Since late 1950s, the random matrix theory was developed in the
   studies of spectra emissions from heavy nuclei by Wigner,
   Dyson, Mehta and many others. \cite{Mehta1991}  So far it has been applied to a large
   variety of fields in physics, mathematics and much more.\cite{Guhr1998, Akemann2011}
   A lot of experimental studies  in real systems also have been done with the RMT, such
   as in mesoscopic systems  and quantum chaos.  
   The RMT has also been applied in various  biological systems including protein-protein interaction networks,
   and the co-expressing gene networks in many organisms.

   There are three categories of RMT depending on their symmetries,
   the Gaussian orthogonal ensembles
   (GOE), the Gaussian unitary ensembles (GUE), and the Gaussian
   symplectic ensembles (GSE).  In the studies of spectra analysis of heavy nuclei, for 
   example, energy levels (eigenvalues) of the unitary Hamiltonian matrices are investigated. 
   The symmetry of the matrix is determined according to the general properties of the systems to be investigated. 
   
   In the limit of large matrix size : $N \rightarrow \infty $ , the distribution of spacings of adjacent eigenvalues 
   (NNL spacings) $P(s)$ becomes a universal function.  Here the term "universal" means that 
   the distribution is independent of any detail of the systems under study and is only 
   affected by its symmetry.  For the above three symmetry groups, $P(s)$ are written together as,   
\begin{equation}\label{eq:pofs}
\begin{array}{ll}
 P(s) &= a_\beta s^\beta  \exp \left( -b_\beta
		    s^2 \right )\cr
a_\beta  &=  \frac{2\Gamma \left((2+\beta)/2)\right)^{\beta+1}}{\Gamma\left((1+\beta)/2\right)^{\beta+2}} , \quad
b_\beta =  \left( \frac{\Gamma((2+\beta)/2)}{\Gamma((1+ \beta)/2)}\right)^2, \cr
\end{array}
\end{equation}
where $s$ is the level spacing rescaled by the mean spacing $D$.  $\Gamma(x)$ is the Gamma
function. The $\beta$ is $1$ in GOE, $2$ in
GUE and $4$ in GSE case, respectively.  

In the GOE case ($\beta = 1$),  the constants become $a_1 = \pi/2$ and $b_1 = \pi/4$,  thus
\begin{equation} \label{eq:wigner}
P(s) = \frac{\pi  s}{2} \exp \left ( -\frac{\pi  s^2}{4} \right ).
\end{equation}
It is called the Wigner distribution.
The Wigner distribution of NNL spacings infers that the eigenvalues have mutual
correlations and repel each other.  It is obvious from the small $s$ behavior where $P(0) = 0$.  
In the opposite case where the eigenvalues
have no correlation and are randomly distributed,  $P(s)$ becomes
\begin{equation}\label{eq:poisson}
P(s) = \exp ( - s).
\end{equation}
This is known as the Poisson distribution in RMT.  We note that it is called the exponential distribution in statistics.  
In many experimental studies including numerical Monte-Carlo simulations, the Wigner distribution of $P(s)$ have been widely observed.
Since the matrix size $N$ is finite in the actual system to be analyzed, we have to apply a method called unfolding which is described explicitly in section 2.4 below.  

\subsection{The interaction matrices for gene networks}
\label{sec:2.3}
In this study, we investigate  the distributions of NNL spacing $P(s)$ of the gene interaction matrices.
The gene interaction matrix  $M$ is evaluated as follows.  

From TCNG, the gene interaction networks were retrieved.  
Each gene interaction network contains a list of interacting gene pairs.
The directions of the inferred edges are omitted.  The gene interaction matrix elements $M_{ij}$ is given by
\begin{equation}
 M_{ij} = 
\left\{
 \begin{array}{ll}
1 & \quad\mbox{if there is an edge between gene $i$ and gene $j$} \cr
0 & \quad\mbox{otherwise}
\end{array}  
\right.
\end{equation}
The $i$ and $j$ are the gene identification numbers.  
For 256 gene interaction networks
in TCNG, we generated 256 corresponding interaction matrices $M$ and the eigenvalues are obtained 
numerically by diagonalizing $M$. We evaluate $P(s)$ for each of the 256 sets of eigenvalues by averaging over segments
of equal number of eigenvalues. 
We set $ M_{ij} = M_{ji}$, then $M$ becomes a real symmetric matrix.  
The self-interaction is neglected: $M_{ii} =0$.  The matrix size $N$ is about $8,000$ for each gene networks
after gene redundancy is omitted.  Interaction matrices $M$ are called 
adjacency matrices in the graph theory. 

The median of the number of non-zero elements in $M$ is about $80,000$.  The gene
interaction networks in TCNG are identified with network indices (the network IDs) from 1 to 256.
The accession numbers for NCBI GEO entries are also tagged by the network IDs and stored in the database.

The 256 NCBI GEO data selected for the Bayesian network calculations in
TCNG are all human cancer related gene expression experiments.  The 119
of them are on the platform Affymetrix Human Genome U133 Plus 2.0 Array
(GPL570) and the 73 on Affymetrix Human Genome U133A Array (GPL96).
There are data on several other platforms from Agilent Technologies and Illumina
Inc., etc.   The numbers of inferred edges in the networks are widely distributed.
The median is about $38,000$, the minimum is about $13,000$,
and the maximum is about $64,000$.  The edge frequencies (likelihood factors)
take values from $0.2$ to $1.0$.  
In TCNG, the edges that have the edge frequencies smaller than $0.2$ have been eliminated.

We divide the edges by the edge frequencies into four groups and $P(s)$ is calculated for each of edge groups.  We first sort the inferred edges in ascending order of edge frequency, and they were equally divided into 4 groups.  Thus each group contains the equal number of edges.   
Although this subdivision of the network is rather intentional,  we obtain the mean node degree $\bar k = 2 \sim 3$ in each of the sub-networks.  Here the node degree $k$ is the number of non-zero elements in each row of the interaction matrix.  

Classifications and extractions of the data have been done using SQLite, 
and the eigenvalue calculations are done by MATLAB (R2017a). 
For the calculations of $P(s)$, the eigenvalues are rescaled with the method called unfolding.
This procedure is briefly described in the next section. 
 
\subsection{The unfolding}
\label{sec:2.4}
In the random matrix theory, the Wigner surmise is valid under the condition that the eigenvalues are uniformly distributed and the spacing between them are very small. This condition hold when the matrix size $N$ is very large and the consecutive eigenvalues are taken from a region not far from zero.  Since $N$ is finite in our numerical analysis using the real data, the rescaling of the eigenvalues called the unfolding method is applied to discuss the large $N$ behavior.

We select $n$ consecutive eigenvalues $x_1 < x_2  < \cdots <  x_{n-1} <  x_n$ from the $N$ eigenvalues.  The width of the selected range is $\Delta E = x_n - x_1$.  The local mean spacings is $\hat D = \Delta E / n$.  The eigenvalue density function is defined 
\begin{equation}\label{eq:rho}
  \rho (x) = \sum_{i=1}^{n}\delta(x - x_i ),
\end{equation} 
where $\delta(x)$ is the delta function.  
The staircase function $\hat N(x)$ is defined as
\begin{equation}\label{eq:staircase}
  \hat N(x) = \int_{x_1} ^x d z \rho(z).
\end{equation}  
$\hat N(x)$ is the number of eigenvalues in the range between $x_1$ and $x$.

When the matrix size is very large, the total number of eigenvalues $N$ is large.  We assume that the distribution of eigenvalues is dense and uniform in the selected local region $\Delta E$.  So the mean spacing $D$ becomes a constant value and the staircase function $\hat N(x)$ behaves as a linear function. The plot of $\hat N(x)$ becomes a straight line with the slope $1/D$ in this region.  The unfolded eigenvalue
$\xi$ is obtained as
\begin{equation}\label{eq:xi}
   \xi = (x-x_1 )/\hat D = \frac{n (x-x_1 ) }{\Delta E}.
\end{equation}

We technically divide $N$ eigenvalues which are sorted in ascending order to $L$ segments which contain the equal number of elements.  The rescaling of the eigenvalues has been done in each of these segments by evaluating the local average $\hat D$.
In this study, the total number of eigenvalues is $N\sim 8000$ and we divide them into $L=76, 38, 25 $ and $19$ segments  each of which
contains $n=100, 200, 300$ and $400$ eigenvalues for the unfolding.    The eigenvalues which lie in the tails of the density function 
$\rho(x)$ in eq.(\ref{eq:rho}) are eliminated.  The unfolded eigenvalues $\xi$ are used to
evaluate the local probability distribution of NNL spacing $\hat P(s)$ in each segment.  We evaluate $P(s)$ by averaging over $L$ segments
which cover $95\%$ of total eigenvalues.   The averaging over segments has also been applied to investigate the eigenvalue statistics  in several studies. \cite{Casati1985} 

\subsection{The hypothesis test of the empirical level spacing  in each segments of the eigenvalues}
\label{sec:2.4}
We tested the null hypothesis "The $(n - 1)$ NNL spacing data obtained from $n$ eigenvalues originate from the hypothesized (exponential or Wigner) distribution." against the alternative by the one-sample Kolmogorov-Smirnov test in each segment.  We set the significance level $\alpha = 0.05$.  
In the Kolmogorov-Smirnov test, the cumulative distribution function (cdf) of the data is compared with the cdf of the hypothesized statistical distribution, and the maximum value of the difference is set as the test statistic.  We evaluate p-values of the hypothesis test with MATLAB
built-in function "kstest".  The p value of the hypothesis test is the probability of observing a test statistic as extreme as, or more extreme than, the observed value under the null hypothesis. 
When the p value is larger than the significance level $\alpha$, the null hypothesis is not rejected.  Note that the p value of the hypothesis test 
does not indicate the probability that the data will match the hypothesized distribution to be tested. The cdf of the empirical level spacing obtained in each segment is plotted together with cdf of eq. (\ref{eq:wigner}) and eq.(\ref{eq:poisson}).  We note that the sample
size of the hypothesis test is $n-1$ in each segment.  

\section{Results}
\label{sec:3}

\subsection{The distributions of NNL spacings depend on gene network sparseness}
\label{sec:3.1}
    We first show results of the gene networks where numbers of edges are less than the
    median $38,000$ of the 256 gene networks from TCNG.  The probability distribution of NNL spacings
    $P(s)$ is obtained from the network ID:236 (NCBI GEO accession number : 
    GSE8057).  There are $32,124$ inferred edges from the
    expression experiment of ovarian cancer cells 51 samples.  We then categorized the inferred edges into
    four groups by the edge frequency (likelihood) factors : 
    $0.2 - 0.25$, $0.25 - 0.3$, $0.3 - 0.5$ and $0.5 - 1.0$.  
    The edge factors $0.25, 0.3$ and $0.5$ corresponds to the
    25th, 50th and 75th percentiles of the sorted edge factors, respectively.  
    
\begin{figure}
\centerline{\includegraphics[width=\textwidth]{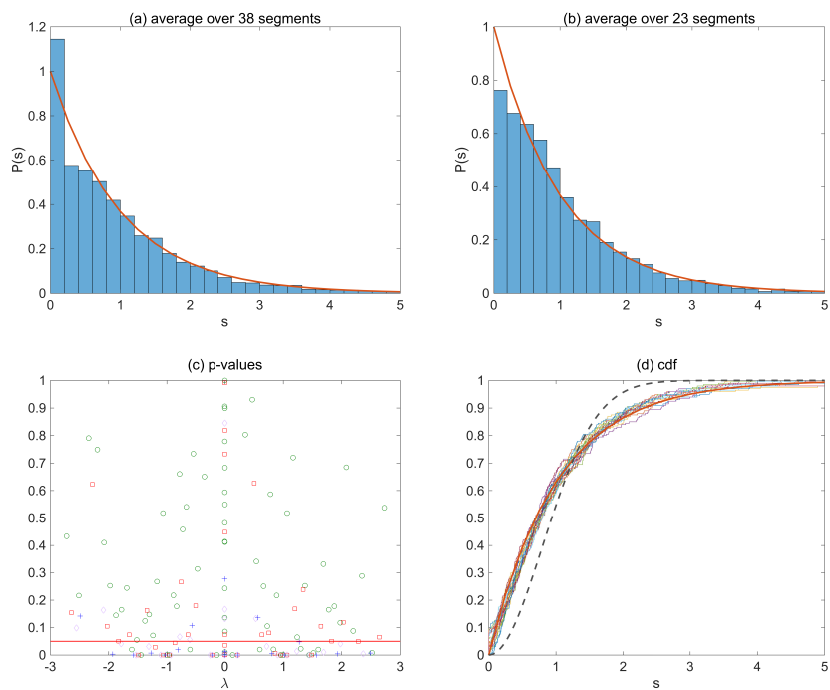}}
\caption{ {\bf The distribution of NNL spacings in a sparse gene interaction network.}
From the gene interaction network in TCNG database [ID:236], the
interaction (adjacency) matrix $M$ is evaluated.  The matrix size is
7,928 which is the number of identical genes after removing duplications.
The total number of edges is 32,124 in this network.  8,031 edges that
are in the highest edge frequency group $(0.5 - 1.0)$ are selected for
the calculation of $M$.  \break 
{\bf (a)} $P(s)$ obtained by averaging over  the $38$ segments which cover 95\% of the total eigenvalues. 
The number of eigenvalues is 200 in each segment.
{\bf (b)} The averaged $P(s)$ over the 23 segments in which the p-values are $p > 0.05$ for a comparison. 
{\bf (c)} The p-values of the one-sample Kolmogorov-Smirnov test for each segments are plotted against the center value of each eigenvalue range.  The red vertical line shows the significance level $\alpha=0.05$.  The number of eigenvalues in each segment is 100(o), 200($\Box$), 300($\Diamond$) and 400(+), respectively. (See also Supplementary Table T1.)
{\bf (d)} The empirical cumulative distribution function (cdf) in the selected 23 segments which contain 200 eigenvalues  each (staircase lines).  The cdf of the distribution in eq.(\ref{eq:poisson}) (bold line) and the cdf of eq.(\ref{eq:wigner}) (dashed line) are shown for comparisons.}
 
\label{fig:01}
\end{figure}

    In Figure \ref{fig:01} the $P(s)$ obtained from the interaction matrix of the
    largest edge frequency group  ($0.5 - 1.0$) is shown.  The bin width is $0.2$.  In Fig. \ref{fig:01}(a), $P(s)$ is obtained by averaging over
    38 segments which contain 200 eigenvalues each.     The distribution $P(s)$ becomes the Poisson distribution.     
    In Fig. \ref{fig:01}(c), we plot the p-values of the one-sample Kolmogorov-Smirnov (KS) test in each segments.  The numbers of eigenvalues
    in each segment is 100(o), 200($\Box$), 300($\Diamond$) and 400(+), respectively.  The p-values in each segments and the 
    boundary eigenvalues of the segments are listed in Supplementary Table T1.  
    We see that while the p-values are largely 
    dependent on the size of the segments, they  seem to become larger in the center part $\xi\sim 0$.   
    
    In Fig. \ref{fig:01}(d), we also show the empirical cumulative distribution 
    function (cdf) in the $23$ segments (200 eigenvalues in each) together with  that of eq.(\ref{eq:poisson}) (bold line) and 
    eq. (\ref{eq:wigner}) (dashed line) for comparison.  
    In these $23$ segments, the p-values of the hypothesis tests are larger than the significance level $\alpha = 0.05$ .  
    On the other hand, the p-values of the KS test for the hypothesized Wigner distribution are less than $0.05$ in all segments.  
    It is obvious from the larger difference between the stepwise lines (the empirical cdf) and the dashed line (the cdf of Wigner distribution) 
    in Fig.\ref{fig:01}(c).  
    We also show the empirical $P(s)$ averaged over the selected $23$ segments in Fig. \ref{fig:01}(b).  
    The convergency to the distribution eq.(\ref{eq:poisson}) in both cases is remarkable.

\begin{figure}
\centerline{\includegraphics[width=0.75\textwidth]{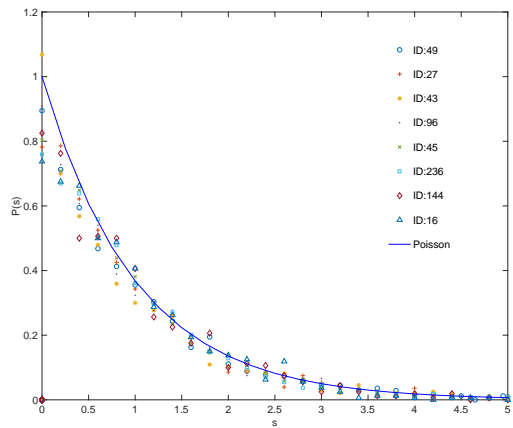}}
\caption{{\bf Distributions of NNL spacings for eight sparse gene interaction networks in TCNG.}
The eigenvalue spacing distributions obey the Poisson distribution in the sparse interaction networks in TCNG database.  
The calculations are done with edges in the highest edge frequency group $(0.5 - 1.0)$ .
The numbers of total edges are,
30 042 (ID:49, GSE13861), 30 214 (ID:27, GSE12391), 30 563 (ID:43, GSE13255), 
30 957 (ID:96, GSE18521), 31 326
(ID:45, GSE13507), 32 124 (ID:236, GSE8057),  34 440 (ID:144, GSE24080) and 34 912 (ID:16, GSE10972),
respectively.  They are chosen from the
sparse network group where the total number of edges are less than
$38,000$ (the median of edge numbers of the 256 gene networks).  
The eigenvalues are unfolded in 38 segments containing $200$ eigenvalues each.
}\label{fig:02}
\end{figure}

    In Figure \ref{fig:02}, we show distributions $P(s)$ for eight gene
    interaction networks in the same class of gene network sparseness, where the
    numbers of the edges are less than $38,000$.   For each of these gene
    interaction networks, the regions of eigenvalues where the local $\hat P(s)$ shows the distribution in eq. (\ref{eq:poisson}) are observed.   
    We apply the one-sample Kolmogorov-Smirnov test in each of $38$ 
    segments and $P(s)$ is evaluated by averaging over the segments which show the p-values  larger than the significance level $\alpha =0.05$.    
        
    We have hardly seen the Wigner distribution of $P(s)$  of eq.(\ref{eq:wigner}) in the sparse group of gene 
    interaction networks for all the quartiles of the edge frequencies.  We also note that the Poisson distribution
    of eigenvalues is independent of the sample size (the number of the expression data) used for the Bayesian network inference.

\begin{figure}
\centerline{\includegraphics[width=\textwidth]{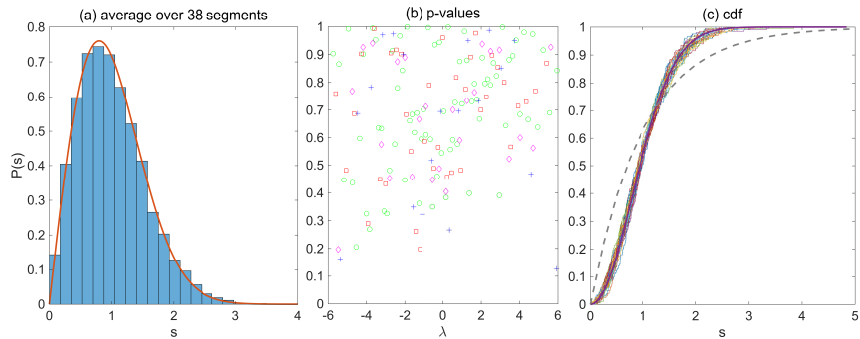}}
\caption{{\bf Distribution of NNL spacings in a dense gene interaction network.}
The NNL spacings distribution for the ID:165 gene interaction
network in TCNG database.  The total edge number is 51,702, and the whole edges are used for the
calculation of the interaction matrix $M$.  The size of the interaction matrix $N=7,996$.  
{\bf (a)} The solid line shows the Wigner distribution eq.(\ref{eq:wigner}).  The $200\times 38$ eigenvalues
are used for the calculation of $P(s)$.  {\bf (b)} The p-values of the one-sample Kolmogorov-Smirnov test for each of $38$ segments.  The red vertical line shows the significance level $\alpha=0.05$.  The number of eigenvalues in each segment is 100(o), 200($\Box$), 300($\Diamond$) and 400(+), respectively.  (See also Supplementary Table T2.)
{\bf (c)} The empirical cumulative distribution function (cdf) in $38$ segments and the cdf for the Wigner distribution in eq.(\ref{eq:wigner}) (bold line).  The dashed line is the cdf of the distribution of eq.(\ref{eq:poisson}) for a comparison. } 
\label{fig:03}
\end{figure}

    In Figure \ref{fig:03}(a), we show the NNL spacings distribution $P(s)$ for a dense gene
    interaction network (ID:165, NCBI GEO accession number : GSE29013), 
    where the number of edges is $51,702$.  This gene interaction network
    inference is done with the gene expression data of 55 samples from
    lung cancer cells.   The interaction matrix is calculated with all edges.
    The Wigner distribution of the local $\hat P(s)$ is observed in all of the $38$ segments independent of the size. 
    The p-values of the one-sample Kolmogorov-Smirnov test for the local distributions $\hat P(s)$ against the
    distribution eq.(\ref{eq:wigner}) is shown in Fig. \ref{fig:03}(b).  We evaluate p-values for four different sizes of the segments.
    The p-values in each segments and the boundary eigenvalues of the segments are listed in Supplementary Table T2.

    In all segments, the p-values are larger than the significance level 0.05 and the alternative hypothesis that 
    “the empirical level spacing does not originate from the hypothesized distribution (Wigner distribution)” is rejected.   
 
\begin{figure}
\centerline{\includegraphics[width=0.75\textwidth]{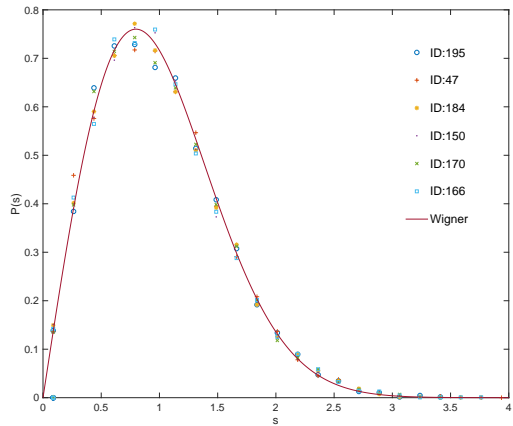}}
\caption{{\bf NNL spacings distributions for six gene interaction networks in TCNG.}
Six gene interaction networks are randomly chosen from the dense group in TCNG database.  The
number of total edges are 50 305 (ID:195, GSE4122), 44 511 (ID:47, GSE13598), 45 789
(ID:184, GSE31547), 45 411 (ID:150, GSE25136), 44 861 (ID:170, GSE29683), and 46 360 (ID:166, GSE2912),
respectively.  }
\label{fig:04}
\end{figure}

    We also show the distributions of NNL spacings for six dense gene interaction networks in Figure \ref{fig:04} 
    altogether.  The ensemble average has been done over the 38 segments which contain 200 eigenvalues each.  
    In all cases, the $P(s)$ show the Wigner distribution.  
       
    The half of the $256$ gene interaction networks in TCNG database have
    more than 38K edges.  The Wigner distribution of
    $P(s)$ for the interaction matrices is widely observed in in the dense network group.  
    We observe the universal  (system independent) behavior of $P(s)$  (Wigner surmise) in the dense gene networks.  
    
    The Wigner distribution of $P(s)$ is also independent of the sample size of the retrieved GEO expression 
    data for the Bayesian network calculations.  We also note that in the gene interaction networks that have less
    than 15,000 edges in TCNG database, $P(s)$ show coincidence neither with the Poisson nor with
    the Wigner distribution for all edge frequency quartiles.

\subsection{Variation of P(s) due to different edge frequencies}
\label{sec:3.2}
    The number of samples that are used for the Bayesian network 
    computations of gene interactions varies from 50 to 559 microarray data samples.  
    The number of samples is, for example, the number of different conditions of the cancer cells or 
    that of patients whose tumor cell is taken in surgery.
    In SiGN-NNSR algorithm, the number of data samples recommended for the Bayesian network
    calculation is more than 100.  
    However,  the computation time of the Bayesian networks also grows heavily as the number of samples
    increases.  It is a characteristic feature of the biological experiments that the number of samples are very limited
    compared to the number of elements (genes) involved.
    
    We also might have a lot of overlooked (false negative) gene interactions due to the experimental noise in each of
    the gene expression data.
    On the other hand, the inference of the gene interactions by the bayesian method can provide lots of 
    false positive edges.  
    In the case where the large part of inferred gene network consists of false positive and false negative interactions (low likelihood edges),  
    the networks may behave as random graphs  where the edge probabilities between any pair of nodes are totally independent each other.    
    We investigate the relationship between the eigenvalue statistics and some graphical properties of subnetworks in each 
    subnetwork grouped by the edge frequencies.

\subsubsection{Graph plots for the gene subnetworks subdivided by the edge frequencies}
All edges of the gene networks are arranged in ascending order with respect to the edge frequency.  
Then we name each edge group (subnetwork) as follows.  
\begin{eqnarray*}
M1 &: &\sim 25 \mbox{ percentiles} \nonumber \\
M2 &: &25 \mbox{ to }  50   \mbox{ percentiles} \nonumber \\
M3 &: & 50 \mbox{ to }  75  \mbox{ percentiles} \nonumber \\
M4 &: &75  \mbox{ to } 100 \mbox{ percentiles} \nonumber\\
ALL &: & \mbox{All edges} \nonumber \\
\end{eqnarray*}

\begin{figure}
\centerline{\includegraphics[width=0.9\textwidth]{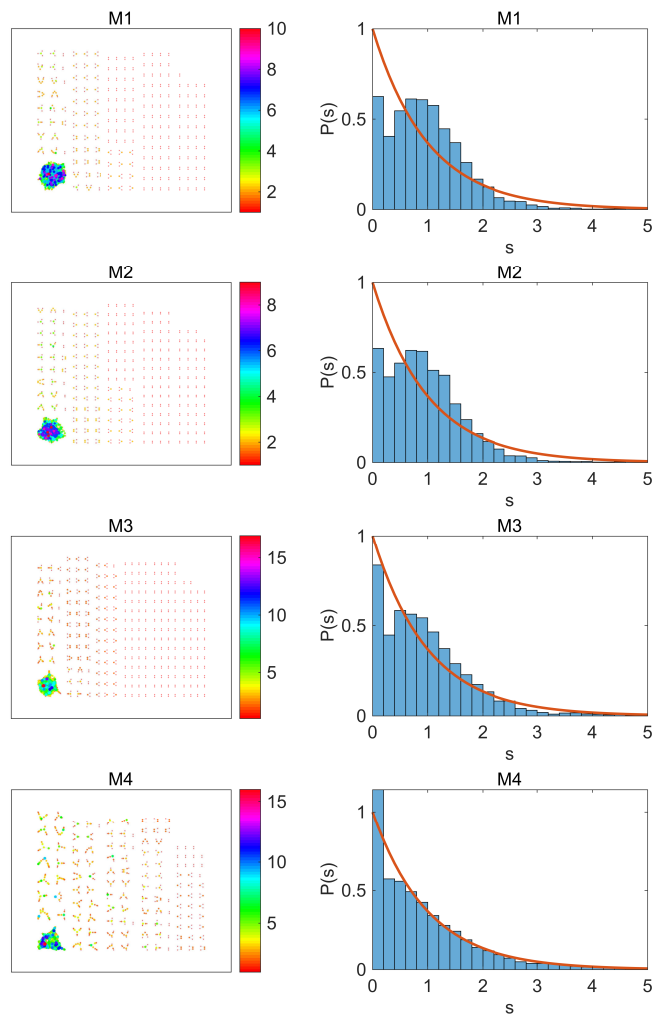}}
\caption{{\bf The gene network subgraph behaviors in the sparse network ID:236}
The color of the node indicates its degree $k$, which is shown in the color bar. 
The size of the node is proportional to $k$.
The ensemble averaging was done over 38 segments which contain 200 eigenvalues each for the evaluation of $P(s)$.}
\label{fig:05}
\end{figure}

In Figure \ref{fig:05},  we show graph plots of the four subnetworks in the ID:236 gene network.   The modularity of the extracted gene network in the M4 edge group is obvious.  We evaluate $P(s)$ for each edge subgroup and show them together in Fig. \ref{fig:05}.  
The Poisson distribution of  level intervals is found in the moderately modular subnetworks M3 and M4.  However, in  the subnetworks M1 and M2, 
the Poisson distribution is lost.  We suggest that it is due to an increase in the proportion of isolated node pairs in these subnetworks.   

\begin{figure}
\centerline{\includegraphics[width=0.9\textwidth]{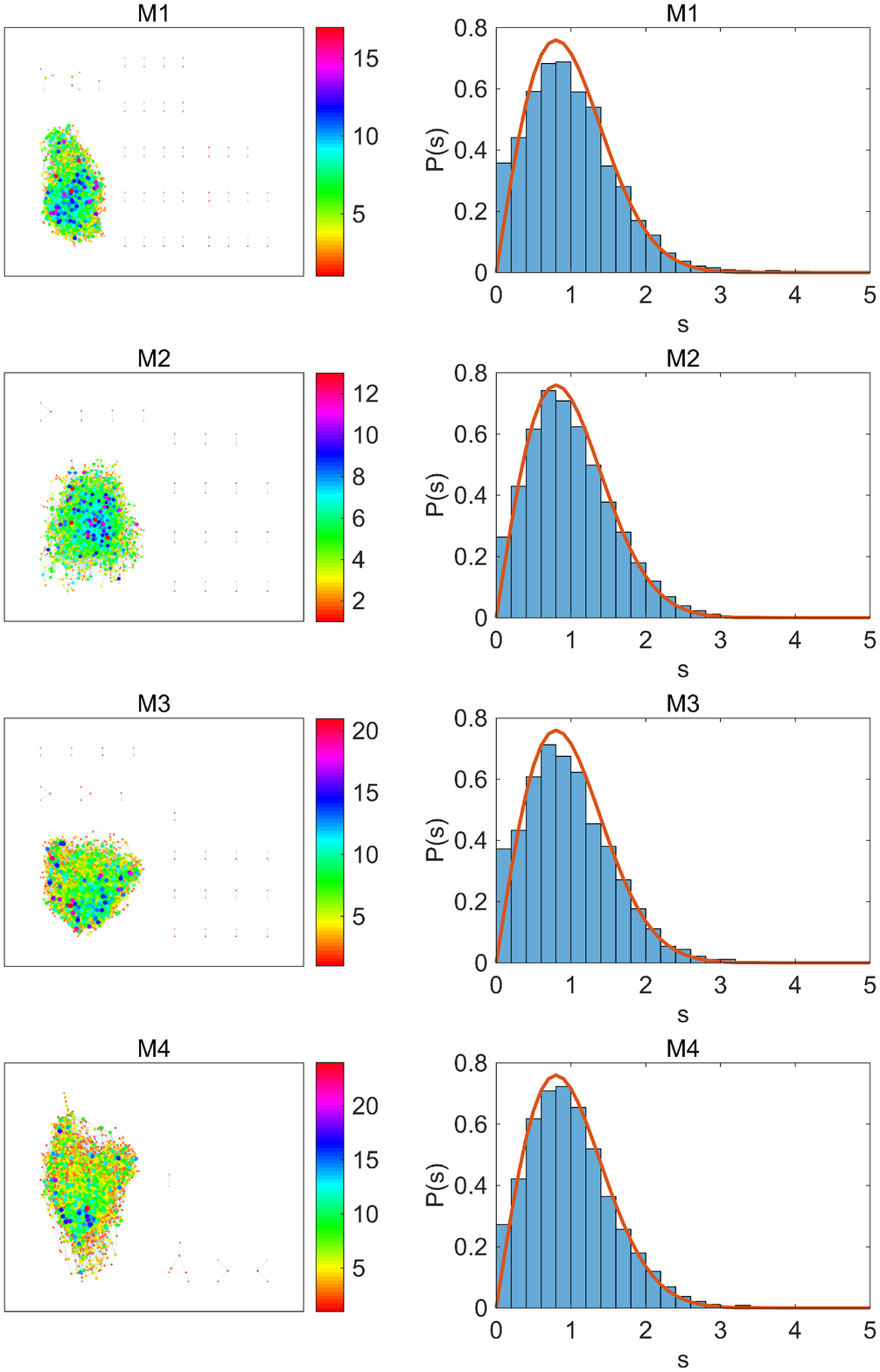}}
\caption{{\bf The gene network subgraph behavior in the dense network ID:165}
The color of the node indicates its degree $k$, which is shown in the color bar. The size of the node is proportional to $k$.
The ensemble averaging was done over 38 segments which contain 200 eigenvalues each for the evaluation of $P(s)$.}
\label{fig:06}
\end{figure}

On the other hand, in the dense gene networks, all subnetworks M1, M2, M3, and M4 show similar "hairball" graphs as shown in Fig.
\ref {fig:06}. 
Wigner distribution of $P(s)$ is seen in all subnetworks regardless of the edge frequencies.
The universality of the Wigner's surmise in the large $ N $ limit is confirmed in the dense gene networks.

The graph plots of all edges of the gene networks ID:236 and ID:165 are shown in Supplementary Figures S1 and S2, respectively.    

\subsubsection{The distribution of the node degree}
\begin{figure}
\centerline{\includegraphics[width=\textwidth]{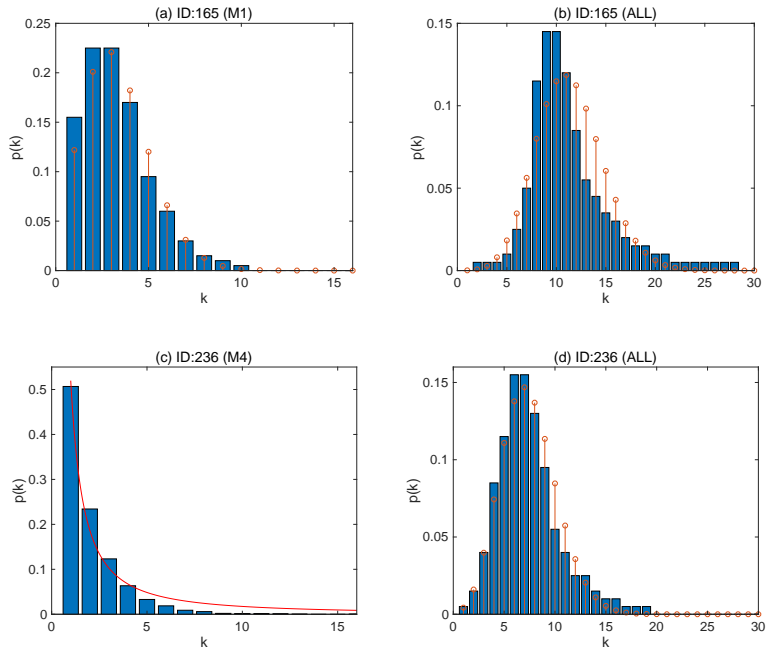}}
\caption{{\bf The probability distribution of the node degree $p(k)$}
The distributions of the node degrees $p(k)$  in two gene networks ID:165 and ID:236.  
In (a), (b,) and (d), the stem plot shows the Poisson distribution given by eq. (\ref{eq:pofk}) with the parameter $\lambda =$ mean($k$) for a comparison.  In (c),  the red line is a function $p(k) \sim k^{-\gamma}$ where $\gamma = 1.5$.  
 }\label{fig:07}
\end{figure}

We plot in Figure \ref{fig:07} the distribution of the node degree $p(k)$ in  four edge groups M1 and ALL in the dense gene network (ID:165), M4 and ALL in the sparse gene network (ID:236), respectively.
Also in Table \ref{tab:01}, the mean degree $\bar k$ and the maximum of $k$   in each subnetworks are listed.

As seen from Tab.\ref{tab:01}, the difference in the number of nodes in the subnetworks M3 and M4 is approximately $1300$  in the sparse network ID:236, which is almost $16\%$ of the total number of edges.    The lost nodes are the nodes whose  edges have only the frequencies 
within the range of the M4 group.  The node which has the largest max($k$) is called the hub.  The smaller values of max($k$) in the M1 and M2 suggest the main hub node is lost in these subgroups.  The lost hub nodes may have edges with larger frequencies only.  
Therefore, it is expected that the major characteristics of the original gene network have been lost in the subnetworks M1 and M2. 

In the case of the random graph, the distribution of the node degrees $p(k)$ is written as
\begin{equation}
p(k) = {}_{n-1}C_k q^k (1 -q)^{n-1-k}, 
\end{equation}
where $n$ is the number of nodes and $q$ is the edge probability.
In the $n\rightarrow \infty$  limit, we take $q\rightarrow 0$ while keeping $nq \rightarrow \lambda$ as finite,  
then $p(k)$ becomes the Poisson distribution.
\begin{equation}
p(k) = e ^{-\lambda} \frac{\lambda^k}{k!} , \quad \{k =1, 2, 3, \cdots \} .
\label{eq:pofk}
\end{equation} 
It is a discrete distribution where the mean and the variance are both given by $\lambda$.
 
In Fig. \ref{fig:07}(a),(b) and (d), we also show the $p(k)$ fitted  by the Poisson distribution given by eq.(\ref{eq:pofk}) with 
$\lambda = {\rm mean}(k)$by a stem plot.  
Compared to the ALL edge group, the Poisson distribution of $p(k)$ can be seen in the M1  
edge groups of the dense gene network ID:165.  (See also Supplementary Figures S3 and S4).   
Although the gene network is highly connected and clustered,  it might be possible that the low likelihood edge subgroups 
are the random graphs.

On the other hand, in Fig. \ref{fig:07}(c) , we see the typical behavior of the scale-free networks\cite{albert2002} $p(k)\propto k^{-\gamma}$ (where $\gamma = 1.5$ is a fitting factor) in the sparse M4 edge group.

\section{Discussion}
\label{discuss}

We have studied the gene interaction networks which are computationally obtained from gene expression experiments of various human cancer cells.  
 We summarize the results in Table \ref{tab:02}.  
    The universal Wigner distribution of the nearest neighbor level spacing $P(s)$ is observed in the dense gene  interaction networks.  On the other hand, in the sparse gene networks, the Poisson (exponential) distribution of $P(s)$  is obtained in the M4 subnetwork.  The threshold of edge number $E$ is about $38,000$, where the change between the dense and the sparse behaviors  of $P(s)$ occurs. 

    The distribution $P(s)$ in the large $N$ limit was obtained by the averaging over segments of equal number of eigenvalues.   The eigenvalues are unfolded in each segment covering $95\%$ of the total  eigenvalues of the interaction matrices excluding the tail part of the
    the density function $\rho(x)$.    
    
    It is significant that the Wigner distribution of $P (s)$ is widely observed in the dense gene interaction networks regardless of the edge frequency factors.  
Wigner distribution of $P(s)$ is also independent of the experimental details of the original expression data used for the inference of the gene networks.  
These results suggest the universality of the Wigner surmise in the gene networks when the matrix size $N$ (the number of nodes) is large.
    
    In the sparse random matrices, the number of nonzero elements in each row (the degree $k$ of each node) takes a finite value in the thermodynamic limit. In the so-called scale-free networks and in the biological networks, the interaction matrix often becomes a sparse matrix. In the gene networks studied in this paper, the interaction matrix can be considered the  sparse random matrix since the mean degree $k$ is around $10$ for  the 8000 nodes even in the dense gene network group ($E > 38,000$). 
    It has been shown both by the replica method and by the super-symmetry method that the eigenvalue distribution of the sparse random matrix has a special behavior at the center part and the tails of the Wigner's semi-circle. \cite{Mirlin1991, farkas2001, semerjian2002, nagao2008}
    From the Supplementary Table 2,  in the 400 eigenvalue segments,  the p-values of the hypothesized Wigner distribution are small both in the vicinity of $x = 0$ and in the large $|x|$ segments.  We thus consider that the eigenvalue statistics of the sparse random
    matrices may have been observed in the "dense" group ($E>38,000$) of the gene interaction networks.  
        
    We also found appropriate subdivision of the network results in the Poisson (exponential) distribution of the level intervals $P(s)$.  
    In the previous studies of gene expression experiments with random matrix method, similar changes of $P(s)$ behaviors have been observed by extracting edges with correlation factors or by the deconstruction of simulated subnetworks.  \cite{Luo2007, jalan2009} 
   We also found that as the network is divided into subnetworks by the edge likelihood,  the gene networks show the modular behaviors.
      We might say that the sparse gene networks in which the $P(s)$ show the Poisson distribution exhibit the decoupling nature of gene  interactions in cancer cells.  
      
   We have also investigated the relation between the distribution of node degree $p(k)$ and the edge likelihood factors (the edge frequencies) in the Bayesian network inferences of gene interactions.  
   When the number of inferred edges is $E>38,000$,  the low frequency (small likelihood) edge groups seem to show the random graph behavior where edge probability $p$ is totally independent each other.  
   On the other hand, in the largest likelihood edge subgroups of the sparse gene network ($E<38,000$),  we observe the distribution of the node degrees behaves as $p(k)\sim k^{-\gamma}$, which is  called as the scale-free behavior.
              
   In this study, we have totally omitted directions of the edges which  may infer gene regulatory relations.
   The discussion whether the Wigner distribution of $P(s)$ is observed also in the directed gene network is very important, since the 
   directionality of the molecular interactions plays an important role in the cell behaviors.  
   We will check whether the Wigner surmise is confirmed also in the directed gene networks in a forthcoming study, 
   and discuss higher order correlations of eigenvalues in the gene networks.    

\section*{Acknowledgements}
The author would like to thank Prof. Shinobu Hikami for valuable comments.

\begin{table}
\caption{The graph statisitics of the subnetworks.  $k$ is the node degree.  }
 \centering
  \begin{tabular}{ccccc}
  \multicolumn{5}{c}{ID:236} \\
    \hline \hline
Edge Group & Number of Nodes &   The  edge frequency & mean $k$ &   max $k$  \\  
 \hline  \hline 
    M1   &     	6849 &         	 0.20  - 0.25  	& 2.33         &           10      \\      
    M2   &           6867 &   		 0.25 - 0.30  	& 2.3271      &           9      \\
    M3    &           6623 &	  	 0.30 - 0.53	&  2.2141       &          17    	\\     
    M4     &          7928  &		 0.53 - 1.0 	& 2.026         &          16     	\\   \hline
    ALL      &           7966  &		0.2 - 1.0	&  7.5368            &      37      \\
   \hline \hline 
     \multicolumn{5}{c}{} \\
   \multicolumn{5}{c}{} \\
    \multicolumn{5}{c}{ID:165} \\

    \hline \hline
Edge Group &   Number of Nodes  & The edge frequency &  mean $k$ &   max $k$   \\  
 \hline  \hline 
    M1   &              7598	&	0.20 - 0.26	&3.3469        &           17      \\      
    M2   &              7715	&	0.26 -  0.30 	& 3.318      	&           13     \\
    M3    &             7721	&	0.30 - 0.50 	& 3.2162       	&            21       	\\     
    M4     &            7993	&	0.50 - 1.0 & 3.2041       	&             24    	\\   \hline
    ALL      &            7996	&	0.2 - 1.0	 &11.6286       &      	55      \\
\hline \hline 
 \end{tabular}
 \label{tab:01}
\end{table}

\begin{table}
  \caption{Summary of the results shown in Figure 1 - 4.    $E$ is  the number of edges. }
  \centering
  \begin{tabular}{c|c|c|c}
    \hline \hline
    Network ID & Network density &   The edge frequency group  & $P(s)$ \\
    \hline 
    16, 27,43, 45,49,  & Sparse &
        M4 & Poisson distribution
 \\
    96, 144,236 &  $ E  < 38000$  &  & \\
    \hline
    47, 150, 165, 166,   & Dense &
    M1, M2, M3, M4 and ALL & 
        Wigner distribution 
\\
    170,184, 195 & $E  >  38000$  &  & \\
    \hline \hline

  \end{tabular}
  \label{tab:02}
\end{table}

\end{document}